\documentclass{elsart}
\usepackage{graphics}
\usepackage{graphicx}
\usepackage{amssymb}
\usepackage{amsmath}

\begin{document}

\begin{frontmatter}

\title{Distribution of Korean Family Names}
\author{Beom Jun Kim\corauthref{cor1}}
\ead{beomjun@ajou.ac.kr}
\author{and Seong Min Park}
\corauth[cor1]{Corresponding author}
\address{Department of Molecular Science
     and Technology, Ajou University, Suwon 442-749, Korea}

\begin{abstract}
The family name distribution in Korea is investigated
in comparison with previous studies in other countries.
In Korea, both the family name and its birthplace, where the
ancestor of the family originated, are commonly used
to distinguish one family name from the others.
The family name distributions with and without the information
of the regional origins are analyzed by using different
data sets of various sizes, and compared with previous
studies performed in other countries.  
The growth rate of the family is empirically obtained.
Contrary to commonly used assumptions, the growth rate 
is found to be higher for the smaller family.

\end{abstract}

\begin{keyword}
family name distribution \sep Zipf's law \sep growth rate \sep population

\PACS 89.65.-s \sep 89.65.Cd \sep 89.20.-a \sep 87.23.Ge

\end{keyword}
\end{frontmatter}

\section{Introduction}

Different from other countries, Korea is known to have relatively small
number of family names. For example, the nationwide survey in 2000 has shown
that  the population of the family name Kim is about 10 millions, which
means that about one among four Koreans is Kim. In general, the family
names have been devised to distinguish one family from others.
Accordingly, if the number of family names is too small, or family
sizes are too big, the distinction with other families becomes very 
inefficient. From this reasoning, it is natural that Koreans have 
developed additional way to give a kind of sub-system by using
the information of the birthplace of the family name, i.e., where
the ancestor of the family came from.
For example, one of the authors of the present paper has the family
name Kim from Gimhae, a region in southern Korea.
The family name Kim in 2000 survey has been found to have
348 different regional origins.
Almost all Koreans know the regional origins of their family names from
their births and the state keeps this information in the population
registration.

In Korean culture, having an unfamiliar family name is not common at all,
and most people hesitate to invent new names. 
For 15 years from 1985 to 2000 (Table~\ref{mytable}), 
the number $N_f$ of family names in Korea
increased only by 11 ($11/277  \approx 4\%$ increase). 
In contrast, the number $N_r$ of the family names with
regional origins (e.g., Kim from Gimhae is taken as different
than Kim from Gyeongju), is increased in the same
period by 829 ($829/3359= 25\%$ increase) (see Table~\ref{mytable}).
Consequently, this shows that most Koreans consider inventing 
new family name as a taboo, while branching out by using new 
regional origins (but with the same family name) is totally 
acceptable in Korean culture.

In this work, we study the distributions of the Korean family names
with and without regional origins. From the above observations, 
one expects that the distributions look very different with and 
without the regional origins, and that the latter distribution is 
similar to other countries, where having new family name is not a taboo
as in Korea.

\begin{table}
\caption{
Date sets:  Korea00 and Korea85 are for the total populations in Korea 
in the years 2000 and 1985~\cite{nso}.
Seongnam98, Osan04, and Hwaseong04 are extracted from telephone books, 
and Ajou03 is obtained from the list of registered students in Ajou University. 
$N_r$ and $N_f$ are numbers of family names with and without
the regional origins, and $N$ is the population.}

\begin{tabular}{cccc}
\hline\hline
Data set    &  $N$  &    $N_f$   & $N_r$  \\ \hline
Korea00 &  45,985,289        &    288     &  4,188 \\
Korea85 &  40,419,652        &    277     &  3,359 \\
Seongnam98 &   248,460          &    161     &   - \\
Osan04   &   19,632           &    114     &   - \\
Ajou03   &   9,802            &    109     &   - \\
Hwaseong04  &   3,952            &     87     &   - \\
\hline\hline
\end{tabular}
\label{mytable}
\end{table}
The data sets analyzed in this work are as follows
(Table~\ref{mytable}): Korea85 and Korea00
are for the total population of Korea in the years 1985 and 2000, which
were downloaded from Korean National Statistical Office~\cite{nso}.
The set Ajou03 is from the list of registered students in Ajou University 
in 2003, and the other sets Seongnam98, Osan04, and Hwaseong04 are extracted
from telephone books published in the corresponding cities
at years 1998, 2004, and 2004, respectively. We have
information of the regional origins only in Korea85 and Korea00.
Throughout the paper, $N$ is the size of populations, $N_r$ and $N_f$
are the numbers of family names with and without regional origins, 
respectively. For the $k$th family name (when the family name
is arranged from the biggest family to smallest one in descending
order), $f(k)$ is the number of people who have that name, leading
to $\sum_{k=1}^{N_f} f(k) = N$. We also use the integrated
probability distribution function $P_{\rm int}(n)$, which 
measures the proportion of the families with the size greater
than $n$.

The paper is organized as follows: In Sec.~\ref{sec:name}, which
contains the main results of the present work, we
study various aspects of family name distributions without
regional origins. The relationships among the three basic quantities, 
$N_f(N)$, $f(k)$, and $P_{\rm int}(n)$ are analytically found
and compared with empirical results. The distributions for
various sizes of populations in Table~\ref{mytable} 
are  compared with previous results in other countries.
The empirically obtained growth rate as a function of
the family size is also discussed.
In Sec.~\ref{sec:regionname}, we perform analysis on the
distributions of family names with regional origin (e.g., Kim's
from different regional origins are considered to be
different names). 
Finally Sec.~\ref{sec:conculsion}
is devoted to discussions and conclusions.

\begin{figure}
\centering{\resizebox*{0.78\textwidth}{!}{\includegraphics{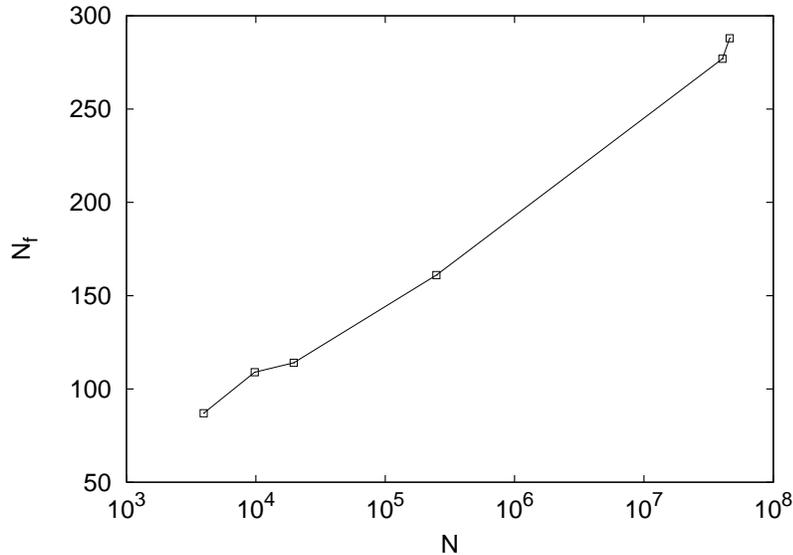}}}
\caption{
The number $N_f$ of family names without regional origins versus
the population $N$. As the population is increased, $N_f$ is
shown to increase logarithmically. Note that only the horizontal axis
is in log scale.
Data are from Table~\ref{mytable} and the line is only guide to eyes. 
}
\label{fig:NfN}
\end{figure} 

\section{Family Name Distribution}
\label{sec:name}
\subsection{Relations: $N_f(N)$, $f(k)$, and $P_{\rm int}(n)$}
\label{subsec:relations}

We first present in Fig.~\ref{fig:NfN} the number $N_f$ of family names
versus the population. As the population is increased, the number
of family names is shown to increase logarithmically. This observation
is in a sharp contrast to Ref.~\cite{japan}, where $N_f \sim N^{0.65}$
has been observed in Japan, instead of $N_f \sim \ln N$. Due to  the lack
of information, we are not able to study $N_r$ versus $N$.

\begin{figure}
\centering{\resizebox*{0.78\textwidth}{!}{\includegraphics{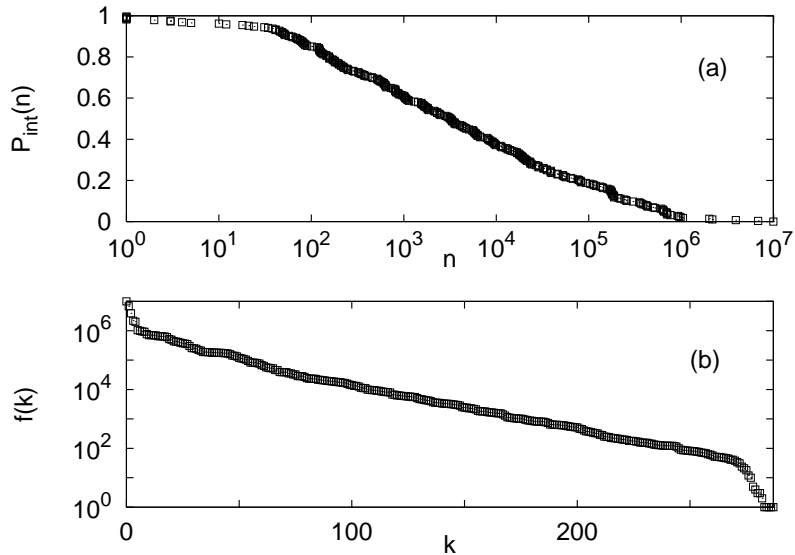}}}
\caption{
(a)Integrated probability function $P_{\rm int}(n)$ versus the family size $n$.
$P_{\rm int}(n)$ measures the ratio of the number of families of the
sizes not less than $n$ to the total number of family names. In a
broad range of $n$, $P_{\rm int}$ is described very well by 
$P_{\rm int}(n) \sim -\ln n$.
(b) The population $f(k)$ as a function of the rank $k$ of the 
family (Zipf's plot).
The biggest family, Kim in this case, has the rank $k=1$. 
In a broad range, $f(k)$ shows exponential decay. 
Both (a) and (b) are obtained from the data set Korea00 
(see Table~\ref{mytable}). 
}
\label{fig:pintfk}
\end{figure} 

In Fig.~\ref{fig:pintfk}(a), the integrated distribution function $P_{\rm int}(n)$,
measuring the proportion of the family names with the populations
greater than or equal to $n$ is displayed. Since every family has at least
one person as its member, one gets $P_{\rm int}(n=1) = 1$. 
As $n$ is increased, $P_{\rm int}(n)$
is a monotonically decreasing function of $n$ until $P_{\rm int} = 0$
is reached
for $n > \max_k f(k) = f(1)$. The logarithmic decay form in
Fig.~\ref{fig:pintfk}(a) is again very much different from the corresponding
results for other countries, where the power-law form 
$P_{\rm int}(n) \sim n^{1-\gamma}$ has been observed with $\gamma  \approx 2$ 
for U.S.A and Brazil~\cite{zanette}, and $\gamma  \approx 1.75$ for 
Japan~\cite{japan}.
The probability
distribution function $P(n)$, which is simply the normalized 
histogram, is easily obtained from the derivative of $P_{\rm int}$
with respect to $n$ from $P_{\rm int}(n) = \int_n^\infty P(n') dn'$,
i.e., $P(n) = -dP_{\rm int}/dn$. In a practical point of view, $P_{\rm int}(n)$
is more convenient than $P(n)$, since some practical issues like the
bin size in $P(n)$ are not needed to be taken care of. 
As will be clearly shown below the same logarithmic form in
Figs.~\ref{fig:NfN} and \ref{fig:pintfk}(a) is not accidental.

Another convenient and frequently
used way of showing the distribution is the so-called Zipf's
plot~\cite{zipf}, where the population $f(k)$ of the family 
name is plotted in terms of the rank $k$.
Figure~\ref{fig:pintfk}(b) is the Zipf's plot of $f(k)$ obtained
from the data set Korea00 (see Table~\ref{mytable}). Clearly 
exhibited is that $f(k)$ decays exponentially with $k$, again
in contrast to Ref.~\cite{japan} where the power-law
decay has been observed. 

The Zipf's plot of $f(k)$
in Fig.~\ref{fig:pintfk}(b) and $P_{\rm int}(n)$ are easily related
since $P_{\rm int}(n)$ is simply the number of family names
of the size greater than $n$: Draw the horizontal line at 
the vertical position $n$ in Fig.~\ref{fig:pintfk}(b) and the value
of $k$ at the crossing point is simply $P_{\rm int}(n)$ multiplied
by $N_f$. In other words, once the functional form $f(k)$ is given,
we obtain the relation
\begin{equation} \label{eq:pint}
P_{\rm int}(n) = f^{-1}(n)/N_f, 
\end{equation} 
connecting the Zipf's plot and the integrated probability
function\footnote{Similar result when both $f(k)$ and $P_{\rm int}(n)$
are of the power-law form has been discussed in Ref.~\cite{japan}.}.
The function $f(k)$ is always a monotonically decreasing function
of $k$ by definition, which confirms the existence of the inverse
function $f^{-1}(n)$.
Via Eq.~(\ref{eq:pint}) the logarithmic form of $P_{\rm int}(n)$ in Fig.~\ref{fig:pintfk}(a)
implies the exponential form of $f(k)$ in Fig.~\ref{fig:pintfk}(b)
and vice versa.

We next establish the simple mathematical relation between 
$P_{\rm int}(n)$ obtained for the total population $N_{\rm tot}$ 
and $N_f(N)$ obtained for various sizes $N$ of population. 
The only one assumption we make is that the set of 
$N$ ($ = x N_{\rm tot}$ with $ 0 < x < 1 )$ individuals is taken 
at random without any bias\footnote{The assumption appears to be very
reasonable in modern societies or in big cities, while it may
not be valid for small rural towns where the family name distribution
is far from the one for the whole country.}.
In other words, the family at the rank
$k$ has $f(k)$ members in total, and thus in the subset of the
size $x N_{\rm tot}$, there are $x f(k)$ members of the same name.
Consequently, once $f(k)$ for the total population is given, 
$N_f$ for the subset of the size $xN_{\rm tot}$ is simply the 
number of family names that has more than one individual,
[$xf(k) > 1$], which leads to
\begin{equation}  \label{eq:Nfx}
N_f (x N_{\rm tot}) = \sum_k \Theta \bigl( f(k) - 1/x \bigr),
\end{equation} 
where $\Theta(y)$ is the Heaviside step function with $\Theta(y) =
1$ for $y > 0$ and $\Theta(y) = 0$ otherwise. From the definition
of $f(k)$, the right-hand side of Eq.~(\ref{eq:Nfx}) equals to
the value of $k$ at the crossing point of the two curves
$f(k)$ and $1/x$, yielding
\begin{equation}  \label{eq:NfN}
N_f (N) = f^{-1}(N_{\rm tot}/N) = N_f(N_{\rm tot})P_{\rm int}(N_{\rm tot}/N),
\end{equation} 
where $N = xN_{\rm tot}$ and the identity in Eq.~(\ref{eq:pint}) have
been used. It should be noted that the relation between the Zipf's plot
of $f(k)$ and the family name distribution function in Eq.~(\ref{eq:pint})
should hold in any case, while the relation between $N_f(N)$
and $P_{\rm int}$ holds only when the assumption of unbiased
random sampling of population is valid. As a specific example,
for the power-law form $P_{\rm int}(n) \sim n^{-a}$, 
$N_f(N) \sim N^a$ is expected. It is not clear whether or not this
relation has been violated in Ref.~\cite{japan}, where
somehow self-contradictory results
$P_{\rm int}(n) \sim n^{-0.75}$ and $N_f(N) \sim N^{0.58}$ have
been concluded from the same unbiased random sampling of population.
However, if the proper size of errorbars are taken, we strongly believe
that the discrepancy should vanish.

\begin{figure}
\centering{\resizebox*{0.78\textwidth}{!}{\includegraphics{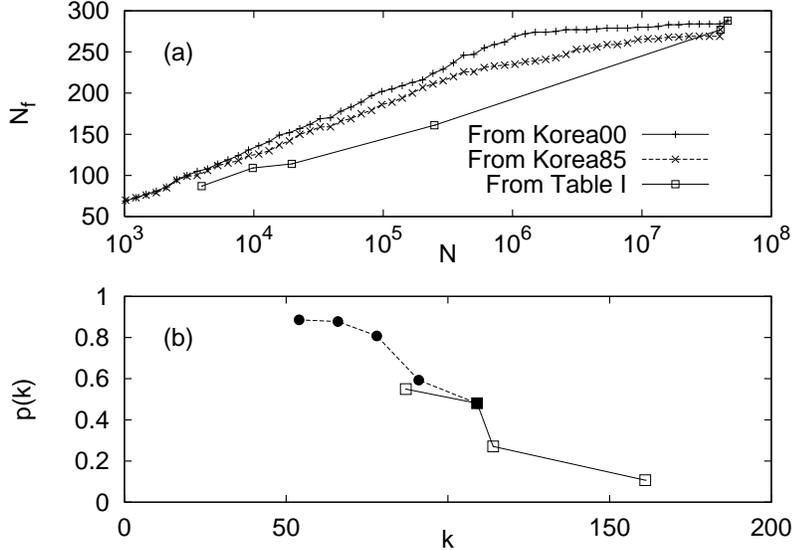}}}
\caption{
(a) The number $N_f$ of family names versus
the population $N$, computed from $P_{\rm int}$ for Korea00 and Korea85
via Eq.~(\ref{eq:NfN}). For comparison, data points in Fig.~\ref{fig:NfN}
are included.
(b) The probability $\Pi(k)$ that an individual of the family name at the rank $k$
is selected by the random sampling of the population. $\Pi(k)$ is
a monotonically decreasing function of $k$, implying that people
with rare names are difficult to be found in urban cities. The points
are obtained by using Eq.~(\ref{eq:pNf}) and values in 
Table~\ref{mytable} for Seongnam98, Osan04, Ajou03, and Hwaseong04.
The data point marked by filled square is for Ajou03 and
is somehow special since it is not based on the residential location
but from the list of students in a university. Points denoted by
the filled circles are obtained from random sampling within Ajou03.
}
\label{fig:NfNpk}
\end{figure} 

In Fig.~\ref{fig:NfNpk}(a), $N_f(N)$ from $P_{\rm int}$ for Korea00
and Korea85 are displayed together with the values in Table~\ref{mytable}.
Although all curves show the logarithmic dependence, i.e., $N_f \sim \ln N$,
$N_f(N)$ from $P_{\rm int}$ computed by using Eq.~(\ref{eq:NfN}) 
shows a systematic difference that it lies higher than the empirical 
results at any $N$. This makes us to conclude that the assumption 
of unbiased random sampling of population is not entirely valid
for data sets in Table~\ref{mytable}.
Nevertheless, the logarithmic dependence persists for all three
different curves in Fig.~\ref{fig:NfNpk}(a), which appears to
justify the unbiased sampling assumption as a reasonable approximation.
The substantial difference in Fig.~\ref{fig:NfNpk}(a) can be interpreted
as follows:  Significant number of families exist in localized area. 
For example, in some villages in rural area, most of people 
in the community have the same family name. Another possibility is
that new names originate from foreigners who just became Koreans 
with new Korean names. The latter case can be probably found only
in the biggest city, Seoul.
Those non-uniformly distributed names in a geographic sense 
are captured in nationwide survey but one cannot find them in the
telephone book in cities we are investigating in this work. 
Accordingly, actual values
of $N_f(N)$ can be different and lie lower than the expected values
from Eq.~(\ref{eq:NfN}) based on the assumption of the unbiased 
random sampling.  

We then elaborate the above explanation further by
introducing  the probability $\Pi(k)$  that the name
of the $k$th rank is chosen. The unbiased random sampling corresponds
to $\Pi(k)=1$. The extension of the above derivation of $N_f(N)$ 
is straightforward: The number of people of the name $k$ in 
the subset of population is now given by $x \Pi(k) f(k)$ and the
$N_f(N)$ is the number of family names satisfying $x \Pi(k) f(k) > 1$.
From the same reasoning as before, $N_f(N)$ with $g(k) \equiv \Pi(k) f(k)$
now reads $N_f(N) = g^{-1}(N_{\rm tot}/N)$.
Consequently, one gets the expression of the probability that
the name at the rank $N_f$ is chosen as
\begin{equation} \label{eq:pNf}
\Pi(N_f) = \frac{1}{f(N_f)} \frac{N_{\rm tot}}{N}.
\end{equation}
We then use the values $N_f$ and $N$ from the empirical values 
in Table~\ref{mytable} and then take $f(N_f)$ from the
Zipf's plot for Korea00 to compute $\Pi(k)$ in Fig.~\ref{fig:NfNpk}(b),
where Seongnam98, Osan04, Ajou03, Hwaseong04 have been used.
The data point (marked as filled square) for Ajou03 is
somehow special since Ajou03 is not based on the residential
location of individuals (Table~\ref{mytable}). It is interesting
to note that $\Pi(k)$ is a decreasing function of $k$, suggesting that
a randomly chosen individual in urban cities is more probable
to have top rank names than for the whole nation. This appears
to be consistent with the expectation that rare names (with lower
ranks) are not distributed uniformly across the whole country.
We also sampled randomly $N$ individuals ($N=500, 1000, 2000, 4000$)
from Ajou03 and count how many names ($N_f$) are found in the set.
We then compute $\Pi(N_f)$ in the same way 
[filled circles in Fig.~\ref{fig:NfNpk}(b)].

\subsection{Zipf's plots of $f(k)$}
\label{subsec:zipf}

\begin{figure}
\centering{\resizebox*{0.78\textwidth}{!}{\includegraphics{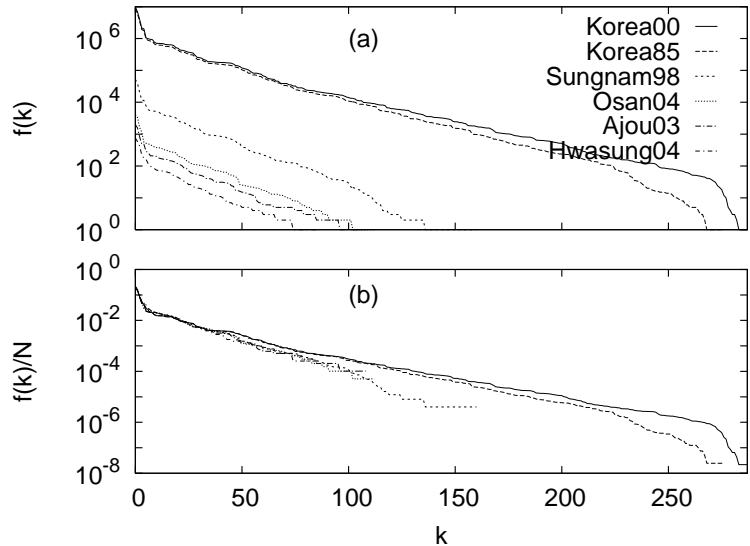}}}
\caption{
(a) The population $f(k)$ as a function of the rank $k$ of the 
family (Zipf's plot) for all data sets (see Table~\ref{mytable}). 
(b) $f(k)/N$ versus $k$ for the same data as in (a).
}
\label{fig:zipfall}
\end{figure} 

In this Section, we investigate the distributions
of family names for various data sets in Table~\ref{mytable}. 
In Fig.~\ref{fig:zipfall}, all data sets in Table~\ref{mytable}
are shown to have qualitatively the same distribution: Zipf's plots 
of $f(k)$ for all data sets show exponential decay form, implying
that the integrated probability distribution follows the
logarithmic form (see discussions in Sec.~\ref{subsec:relations}).
The unanimously found logarithmic form for $P_{\rm int}(n)$
then suggests that the probability distribution $P(n)$, which
measures how many family names have the size $n$, follows the
following form
\begin{equation}
P(n) \sim n^{-\gamma}
\end{equation} 
with $\gamma \approx 1$ in Korea, in a sharp contrast to 
other countries where $\gamma > 1.5$ have been concluded.
This somehow unique  family name distribution in Korea 
can be understood from the birth-death model in 
Ref.~\cite{zanette}, where it has been shown that
$P(n) \sim n^{-1}$ occurs in a stationary state 
if the total population does not grow (or the death rate $\mu = 1$) 
and the rate $\alpha$ of new name generation is very 
small.
The latter condition appears to be fulfilled
in Korea since only very small number of new names were generated 
between 1985 and 2000 (see Table~\ref{mytable}). 
Furthermore, the family names in 
Korea have been introduced to Korean societies 
very long time ago (at least one thousand years ago), and 
therefore one can assume that the Korean family name distribution
is very close to the one at stationary state although the population
is still growing. 
Although new name generation is also very rare in Japan like Korea, 
family names in Japan, on the contrary,  have a short history 
(most names were created about 120 years ago~\cite{japan}). The
qualitatively different family name distribution in Japan
appears to imply that the distribution is still far from
its stationary state as pointed out in Ref.~\cite{japan}.
From this perspective, it is very plausible
that as time proceeds the family name distributions in other
countries also should approach to $P(n) \sim n^{-1}$ eventually,
in the condition that the rate of new name generation becomes
sufficiently small.

\subsection{Growth rate of family size} 
\label{subsec:growth}
When the family name was introduced in human history, it is expected that
the larger the family, the more probable the family's survival, 
since the size of the family was the measure of its strength in many
respects such as the labor power and the number of warriors . 
In other words, the population growth rate of a family  
must have been an increasing function of the size of the family
long time ago.
However, in a modern society, the population growth rate 
is not necessarily an increasing function of the family size
since the size is no longer a matter of life and death.

\begin{figure}
\centering{\resizebox*{0.78\textwidth}{!}{\includegraphics{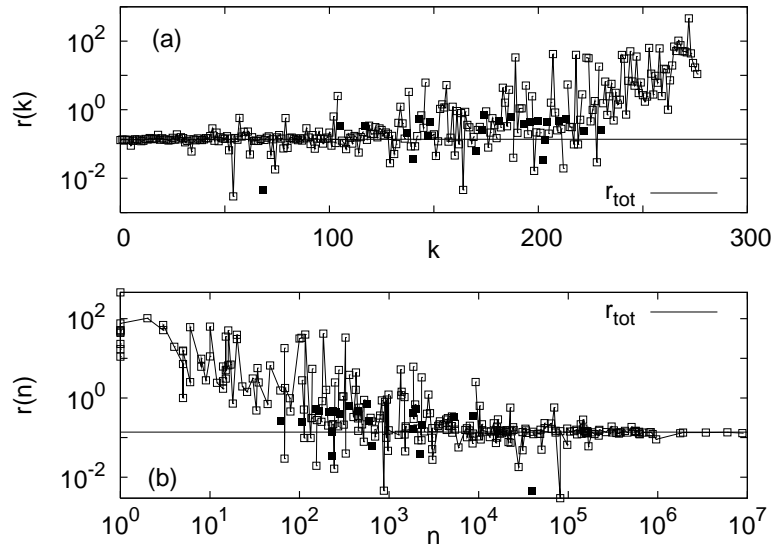}}}
\caption{
The population growth rate $r$ as a function of (a) the rank $k$
of the family and (b) the size $n$ of the family. 
$r$ is obtained from Korea85 and Korea00 from 
$r = [f(k)^{2000} - f(k)^{1985}]/f(k)^{1985}$ with $f(k)^{\rm year}$
being the size of the family of the rank $k$ at a given year.
Sizes of a number of families are found to decrease giving negative
values of $r$. In this case, we plot $-r$ instead and denote those
families as filled squares. The growth rate of total population
$r_{\rm tot} = 0.1377$ is shown (full line) for comparisons.
}
\label{fig:rnf}
\end{figure}

We use two data sets Korea85 and Korea00 to compute the
growth rate of the family size for each family name in 15 years.
The growth rate of the total population is 0.1377 during this
period as shown in Table~\ref{mytable}.
It is found that no name actually disappeared between 1985 and 2000,
but 11 new names were created. In Fig.~\ref{fig:rnf}, we plot
the growth rate $r$ as a function of (a) the rank $k$ and (b) the
size $n$ of the family. There are several very interesting features.
The growth rate of some family is huge. For example,
one family at very low rank had only one individual as its member 
in 1985 while the family size became 462 in 2000, with the growth rate
461 in 15 years. This can never be explained from any biological
reason (no one can have several hundreds of children in 15 years), 
but rather must be a reflection of some sociological or psychological 
forces.  Relatively high growth rates are mostly observed
in families at lower ranks, and $r(k)$ in Fig.~\ref{fig:rnf}(a) is
roughly an increasing function of $k$. 
Such a nonuniform growth rate can be explained from the assumption
that (1) the family name with a high growth rate was recently 
invented and that (2) the family of a smaller size tries harder to 
increase the number of its members by e.g., recruiting new members from 
relatives who still keep the old name. The above assumptions look
reasonable since most people, especially who have small
family sizes, probably want to see their names flourishing, not disappearing
in the future.

In Fig.~\ref{fig:rnf}(b), we display the growth rate $r(n)$ as a function 
of the family size $n$. As expected from Fig.~\ref{fig:rnf}(a), $r(n)$
is a roughly decreasing function of $n$, saturating towards the
total growth rate $r_{\rm tot}$. It is interesting to see that $r(n)$
is significantly higher than $r_{\rm tot}$ when $n \lesssim n_c = 10^3$.
The size $n_c$ can be interpreted as the size scale beyond which 
the nonbiological growth is dominated by the biological growth.
One can also associate $n_c$ with a psychological turning point
in people's mind separating small and big families.

\section{Family Name With Regional Origin} 
\label{sec:regionname}
As an additional way to distinguish families in Korea,
the regional origins of families are simultaneously used in Korea.
In this Section, we regard the family name with different
regional origins as distinctive names, e.g., Kim from Gimhae
and Kim from Gyeongju are considered to be different names,
and study various aspects of the distribution.

\begin{figure}
\centering{\resizebox*{0.78\textwidth}{!}{\includegraphics{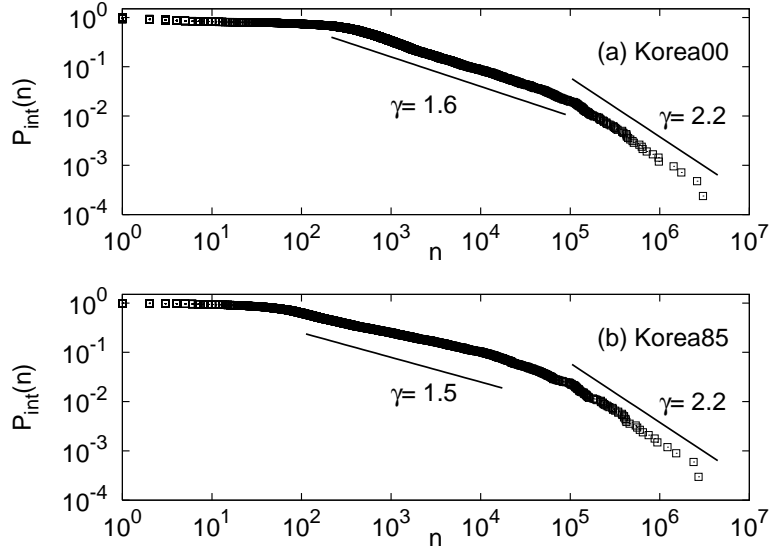}}}
\caption{
Integrated probability function $P_{\rm int}(n)$ versus the family 
size when the name with different regional origins taken as different
for data sets (a) Korea00 and (b) Korea85.
}
\label{fig:pintnr}
\end{figure} 

Figure~\ref{fig:pintnr} shows the integrated probability
function $P_{\rm int}(n)$ measuring the proportion of families 
which have more than $n$ family members. The distribution
in Fig.~\ref{fig:pintnr}, with the broad intermediate range described
by the power-law behavior, is very different from the corresponding
plot in Fig.~\ref{fig:pintfk}. In other words, the family name 
distributions are completely different with and without the 
regional origins. Furthermore, although the distribution 
of the family names without information of regional origins
in Korea is very different from other countries, if the
regional origins are used to distinguish one family from
others the distribution shows the qualitatively similar
behavior to that in other countries.
More specifically, the power-law behavior
$P(n) \sim n^{-\gamma}$  [i.e. $P_{\rm int}(n) \sim n^{1-\gamma}$]
is observed in a broad range of $n$.

\begin{figure}
\centering{\resizebox*{0.78\textwidth}{!}{\includegraphics{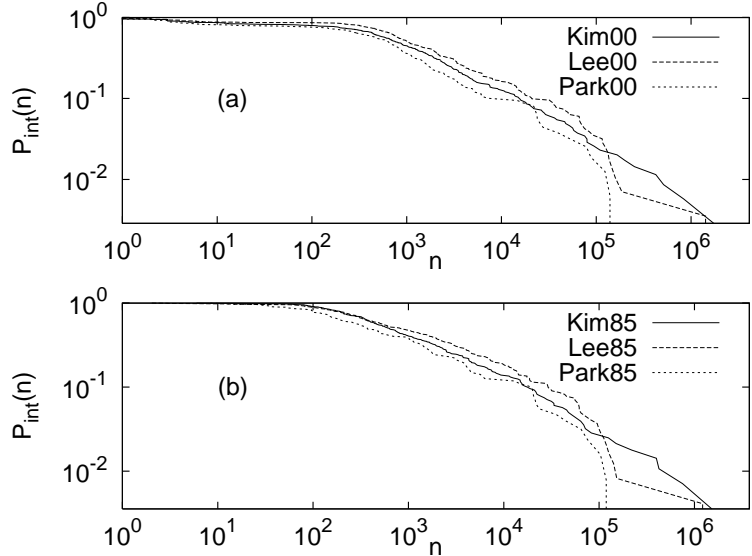}}}
\caption{
$P_{\rm int}(n)$ for subsets with the family name Kim, Lee, and Park
subtracted from (a) Korea00 and (b) Korea85. 
Kim's with different regional origins are considered to be different.
The power-law decay behavior of $P_{\rm int}(n)$ are ubiquitously
seen.
}
\label{fig:nrall}
\end{figure} 
\begin{figure}
\centering{\resizebox*{0.78\textwidth}{!}{\includegraphics{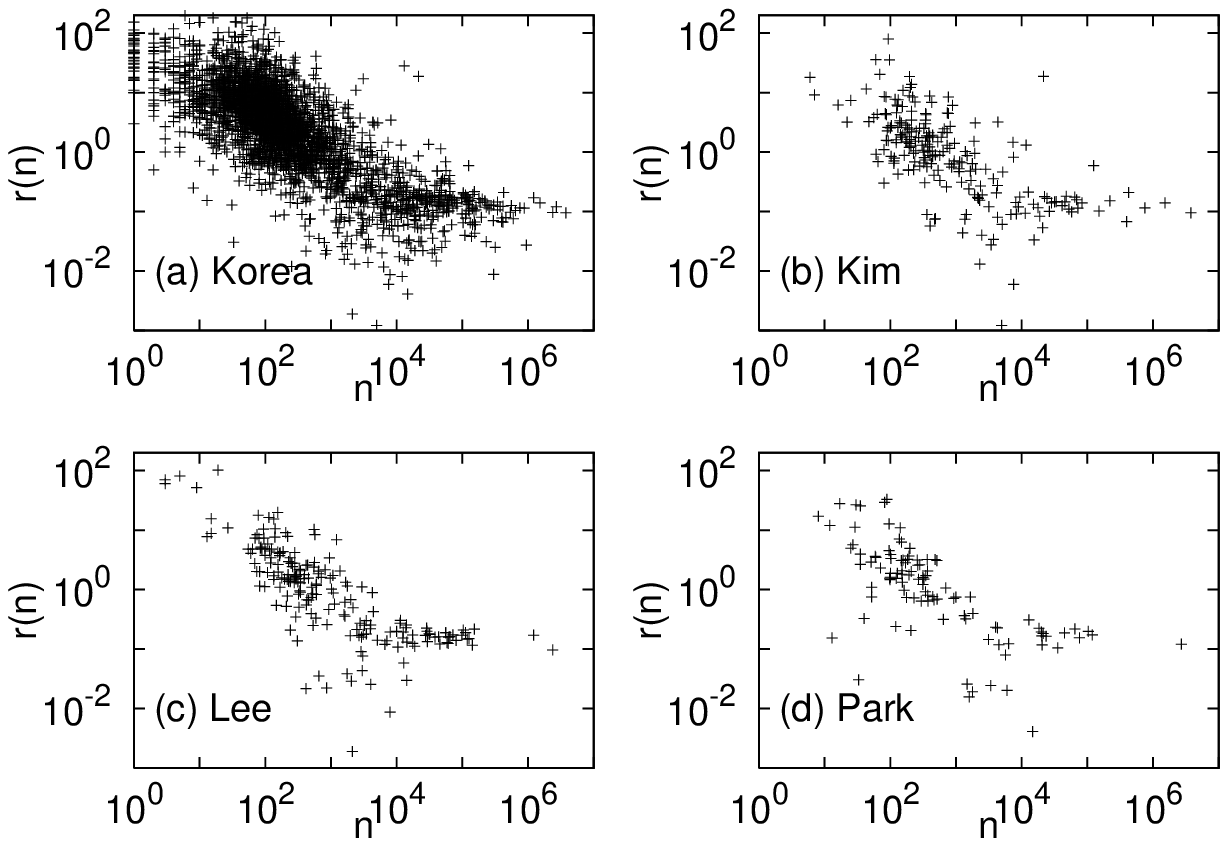}}}
\caption{
Growth rate $r(n)$ as a function of family size $n$
for (a) Korea, (b) Kim, (c) Lee, (d) Park. Identical family
names but with different regional origins are considered as different.
}
\label{fig:rnr}
\end{figure} 

Figure~\ref{fig:nrall} is for the subsets of (a) Korea00 and (b) Korea85.
For example, Kim00 is a subset of Korea00 that contains only the name
Kim but from different regional origins. The three biggest families
Kim, Lee, and Park have 348, 283, and 159 different regional origins
in Korea00, and 281, 244, and 127 in Korea85, implying that new
regional origins have been branching out quite rapidly within 
a family. All curves in Fig.~\ref{fig:nrall} have similar power-law
decay behaviors as in Fig.~\ref{fig:pintnr}.

We finally investigate in Fig.~\ref{fig:rnr} the growth rate $r(n)$
for the family name with regional origin as a function of the
family size $n$ for (a) the whole dataset Korea, and for
three major family names (b) Kim, (c) Lee, and (d) Park; All
show qualitatively the same behaviors as in Fig.~\ref{fig:rnf}
in Sec.~\ref{subsec:growth}, i.e., very high growth rate
for small families, and saturation to average growth rate 
beyond around $n=10^3$. It is interesting to note that
as long as the growth rate $r(n)$ is concerned, the behavior
does not depend much on whether the regional origins are 
taken into account or not. This implies that 
a family with very common family name but with a very
rare regional origin still considers itself as small 
and thus makes tremendous nonbiological efforts to increase
its members.

\section{Conclusion} 
\label{sec:conculsion}

We in this paper have studied the distributions of Korean family names
extracted from various sources: from nationwide surveys in 1985 and 2000,
from telephone books published in three cities, and from the list of
registered student in a university.  Family name with and without
regional origins have been found to show quite different distributions:
Integrated probability distribution is logarithmic without regional
origins (Kim from Gimhae and Kim for Gyeongju are regarded as the same
name), while it is power-law behavior with regional origins (Kim
from Gimhae is considered to be different name than Kim from
Gyeongju). The difference is also reflected in the Zipf's plot
of family size $f(k)$ with the family rank $k$: exponential versus
power-law.

Relation between $P_{\rm int}(n)$ and $f(k)$ have been established,
and $N_f(N)$ (how many family names are found in population $N$) has
been shown to have simple relation with  $f(k)$ (and thus with $P_{\rm int}(n)$)
by using the assumption of random unbiased sampling. 
The empirical observations show systematic deviations from the expected
values, which was then used to compute the probability $\Pi(k)$ that
the family at the rank $k$ is selected if we pick one individual at random.
Interestingly, $\Pi(k)$ has been found to be decreasing function of $k$, implying
that small family names are hard to find in general.

Growth rate $r(n)$ for the family of the size $n$ has been computed
from empirical data (with and without regional origins). All show 
very interesting behaviors: Huge growth rate [$O(10^2)$ in 15 years]
for small families implying nonbiological growth, and saturation
towards average growth rate starting from around $n_c \approx 10^3$,
which has been interpreted as the sociological/psychological separation
point of big and small family.

\section*{Acknowledgements}
We thank S.H. Jeon, S.H. Lee, and D.M. Lee for useful discussions
and acknowledge Ajou University for providing us the list
of registered students. This work has been supported by 
the Korea Science and Engineering Foundation through Grant
No. R14-2002-062-01000-0.

\end{document}